\documentclass[aps,prl,twocolumn,showpacs,superscriptaddress,groupedaddress]{revtex4}  

\usepackage{eurosym}
\usepackage{amssymb,amsmath}
\usepackage{graphicx}

\graphicspath{ {./} }
\begin{document}

\title{Spontaneous multi-keV electron generation from Fermi acceleration and magnetic moment non-adiabaticity}
\author{C. Swanson}
\email{charles.swanson@psatellite.com}
\affiliation{Princeton Satellite Systems, Plainsboro, New Jersey 08536, USA}
\author{S.A. Cohen}
\affiliation{Princeton Plasma Physics Laboratory, Princeton University, Princeton, New Jersey 08543, USA}
\date{\today}

\begin{abstract}
X-ray emission shows the existence of multi-keV electrons in low-temperature, low-power, capacitively-coupled  RF-heated magnetic-mirror plasmas that also contain a warm (300 eV) minority electron population.  Though these warm electrons are initially passing particles,  we suggest that collisionless scattering -- $\mu$ non-conservation in the static vacuum field  -- is responsible for a minority of them to persist in the mirror cell for thousands of transits during which time a fraction are energized to a characteristic temperature of 3 keV, with some electrons reaching energies above 30 keV. A heuristic model of the heating by a Fermi-acceleration-like mechanism is presented, with $\mu$ non-conservation in the static vacuum field as an essential feature.
\end{abstract}

\maketitle

\address{\textit{Princeton Plasma Physics Laboratory, Princeton University,
Princeton, New Jersey 08543, USA}}

Low temperature plasmas, formed at low gas pressure by a low-power radio-frequency (RF) method, are used to study basic plasma phenomena, $\it{e.g.}$, wave propagation and absorption, plasma heating, plasma transport, solar flares, magnetic reconnection, parametric instabilities, turbulence, $\it{etc}$.\cite{malmberg_collisionless_1967,wharton_microwave_1968,yamada_study_1997}, and for numerous practical applications including plasma processing and rocket propulsion.\cite{lieberman_principles_2005,takahashi_high_2017} In these studies and applications, the plasma is assumed to be a cool near-Maxwellian and is investigated with diagnostics whose response and sensitivity are tailored to the low plasma-temperature regime, T$_e <$ 10 eV, though some diagnostics may have an energy range extending to near 100 eV. 

In a previous paper\cite{jandovitz_demonstration_2018} we showed that this assumption is not always justified. We reported the discovery of a minority population of warm (300 eV) electrons created in a low-pressure, low-power, low-temperature, RF-formed plasma. This population was invisible to Langmuir probes;  the electron saturation current of this population was far smaller than the ion saturation current of the bulk. In this paper, we show that another minority population develops and violates the cool Maxwellian assumption (T$_e<$ 10 eV) still more strongly, becoming heated to a characteristic temperature of 3 keV. We explain this by invoking a novel multi-dimensional Fermi-type $2^{nd}$-order heating process. The presence of a keV minority population can strongly affect the properties of these low-temperature basic plasma experiments and processing reactors. 

Other authors\cite{alexeff_oscillations_1968,smullin_generation_1962,demirkanov_plasma_1965,blinov_plasma_1967,arzhannikov_new_1988}, have described the production of higher density (to $10^{13}$ - $10^{15}$ cm$^{-3}$) hot electrons distributions when externally applied high-power (10 kW - 10 GW) pulsed electron beams pass through cool plasmas. The electron heating in these experiments is attributed to electrostatic turbulence along the entire beam, essentially along the entire plasma column. The oscillating electrostatic potential that develops is on the order of the  beam energy, $\it{ca.}$ 10 - 100 kV. 

Multi-keV electrons have also been generated in mirror machines by 1- and 2-frequency electron cyclotron resonance (ECR) heating,\cite{howard_four-dimensional_1986,lichtenberg_velocity_1986,bardet_hot-electron-plasma_1975} a process that improves confinement. The second frequency is typically considered to be required in order to eliminate adiabatic boundaries to electron heating; our heuristic model replaces this second frequency with the natural non-adiabatic mobility of $\mu$ in the static vacuum field.

Herein we report on the spontaneous development of a hot (3 keV) electron minority population in the center cell (CC) of a low-power, sub-ECR, RF-formed mirror-machine plasmas with no high-power electron beam injected. In this experiment, a cm-scale, low current, $\it{ca.}$ 500 eV electron beam spontaneously forms near one mirror coil. This beam generates weak, localized, axial electrostatic field oscillations through two-stream instability. Based on the background gas density, electric and magnetic field characteristics, and power dependences of the hot-electron component’s density and temperature, we attribute the hot electron generation to a combination of several effects, primarily $\mu$-non-conservation and interaction with the electrostatic oscillation at the electron-beam end of the CC. These contribute to a multi-dimensional Fermi-like longitudinal electron heating that is ordinarily assumed to be unsuitable due to heating-induced particle loss and adiabatic barriers to electron energization. 

\begin{figure*}[tbp]
\centering\includegraphics[scale=0.45]{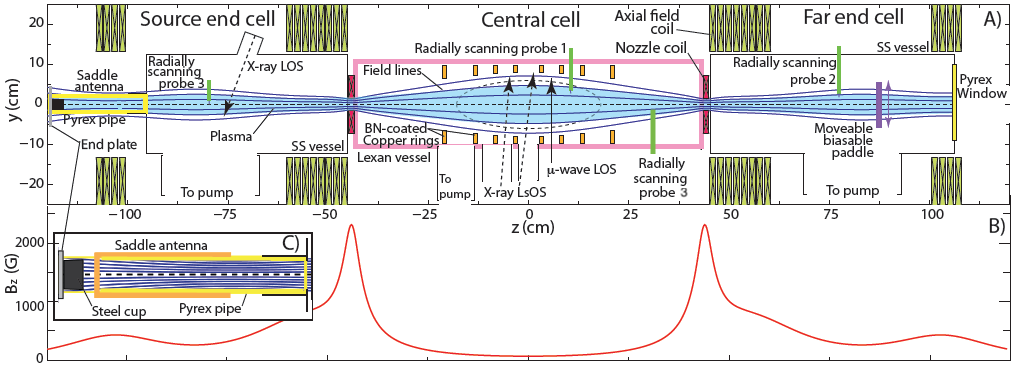}
\caption{Schematic of the apparatus (top); typical axial field strength (bottom); detail of RF antenna region (bottom inset). The primary plasma (light blue) is formed $\it{via}$ RF heating and secondary electron emission in the antenna region of the source end cell\cite{jandovitz_demonstration_2018} and flows into the central cell and far end cell. Three SDD lines-of-sight (LsOS) are shown (dashed arrows).}
\label{fig:schematic}
\end{figure*}

A schematic representation of the apparatus is in Figure \ref{fig:schematic}. It is the same apparatus as used for the PFRC-II experiments,\cite{cohen_first_2012} herein run solely as a low-power magnetic mirror. In these experiments typical central-cell midplane, $\it{i.e.}$, z = 0, magnetic field strengths are 50 - 250 Gauss with a controllable mirror ratio of 10 - 40. The gas fill in the CC is typically 0.1 - 0.2 mTorr of H$_2$ gas. Typical forward power is 300 W. At the neutral density in these experiments, 1 keV electrons have a mean-free path in the CC, based on the total collision cross section, of $\sim$ 10$^4$ cm. 

Because of our interest in higher temperature plasmas, we have been using an electron energy diagnostic, an energy-resolved (pulse-height), rapid-response, silicon-drift X-ray detector (SDD),\cite{noauthor_amptek_nodate} that is sensitive in the range 0.2-100 keV. The SDD has an energy resolution, $\Delta$E/E, where E is the X-ray energy, of 0.03-0.1. The Electron Energy Distribution Function (EEDF) can be extracted from the raw data by a spectral inversion process and an absolute density calibration process  developed for this purpose.\cite{swanson_using_2018} 

Based on information from this detector when viewing the source end cell (SEC), we have previously reported on a population of warm ($T_e$ $\sim$ 300 eV) electrons in the plasma when low-power RF is applied using an external, capacitively-coupled antenna. The cause of these electrons has been shown to be secondary electron emission from antenna-proximate RF-self-biased surfaces in contact with the plasma.\cite{jandovitz_demonstration_2018}

Hot electrons (3 keV, as opposed to warm electrons, 300 eV) are observed when the SDD views across the CC near its midplane. They are observed only when the CC pressure is low, 0.1 - 0.2 mTorr, and not above 0.3 mTorr in the CC.

\begin{figure}[tbp]
\centering\includegraphics[scale=0.42]{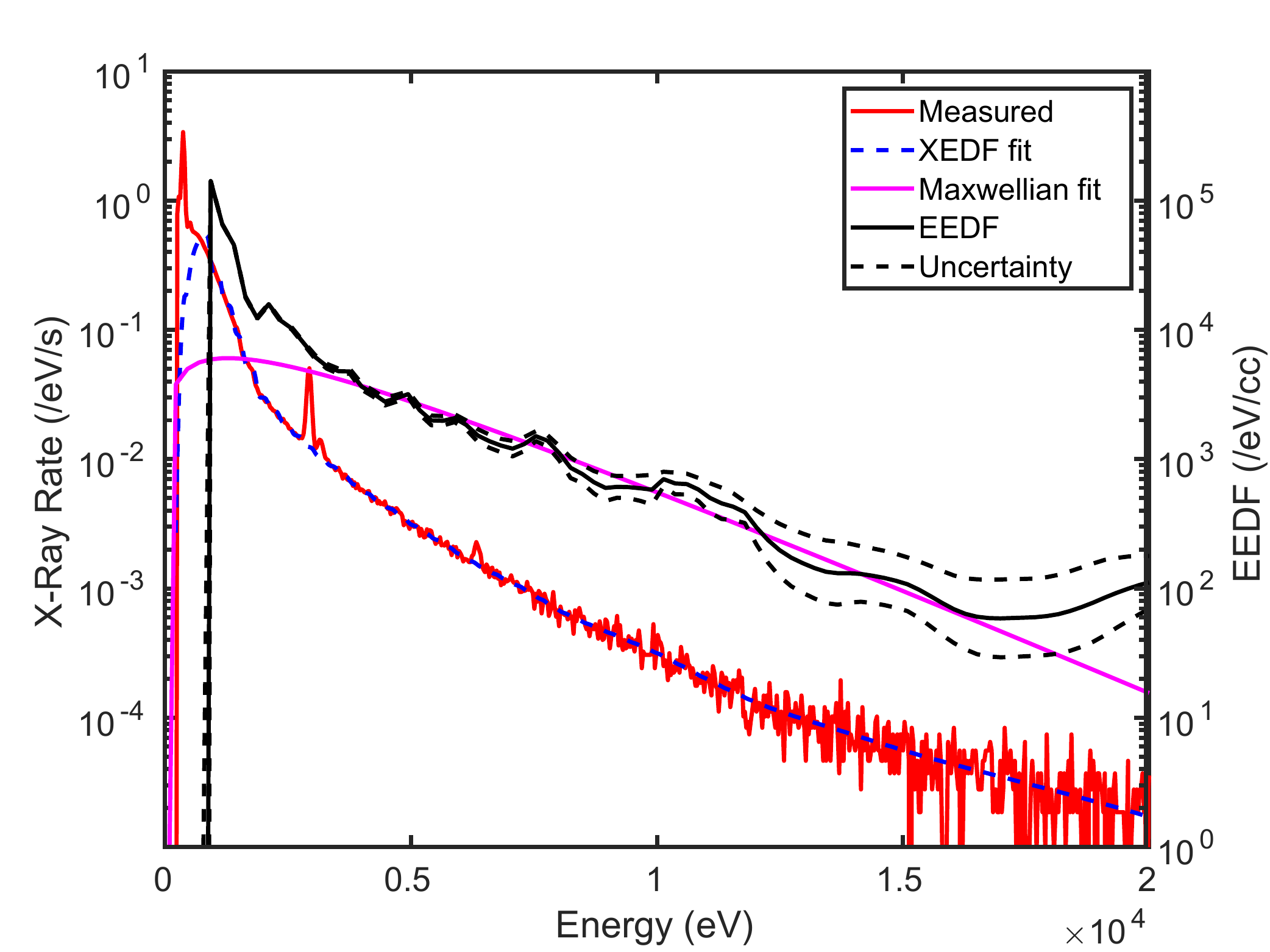}
\caption{Raw X-ray data (red); inverted\cite{swanson_using_2018} EEDF (black solid); 1$\sigma$ uncertainty in EEDF (black dashed); Maxwellian EEDF with $T_e$ = 2.55 keV and $n_e$ = 3.2x10$^7$/cc (magenta). The raw data shows X-rays of energies even out to 30+ keV (not shown) }
\label{fig:EEDF}
\end{figure}

Figure \ref{fig:EEDF} shows an X-ray spectrum measured in the CC and its derived EEDF. (SEC X-ray spectra were simultaneously monitored and showed a 1\% minority population with $T_e$ $\sim$ 300 eV.) The CC spectrum between 3 and 18 keV is well characterized by  $T_e$ = 2.55 keV (hot component); below 2 keV  $T_e$ =500 eV (warm component) is found, about twice that in the SEC under these conditions. Also present are weak spectral lines, mainly of N, Ar and Fe, the latter presumably due to plasma impact on a steel surface near the antenna. 

Radial profiles of the X-ray signals with Ar and Ne fill gases confirmed that wall fluorescence was not the cause of the X-rays.

The bulk electron $T_e$ and $n_e$ in the CC were not typically measured concurrently with the warm and hot electron parameters, but when measured were $T_e \sim 5 \textrm{ eV}, n_e \sim 10^{10} - 10^{11} \textrm{ /cm}^3$. 

\begin{figure}[tbp]
\centering\includegraphics[scale=0.82]{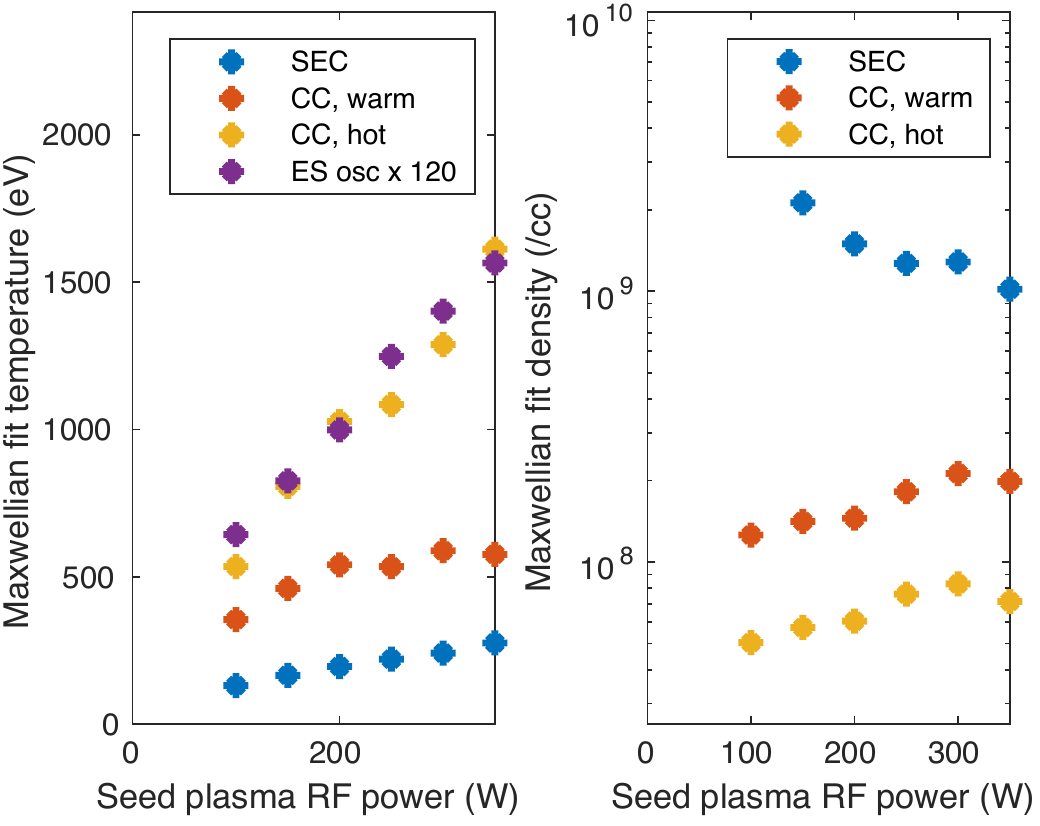}
\caption{Minority electron population $T_e$ and $n_e$ derived from X-ray signals and scaled peak-to-peak electrostatic oscillation (``ES osc'') \textit{vs} the RF forward power. The CC pressure was 0.125 mTorr of H$_2$ gas. Temperature and density uncertainty result from the fit, and are $<10\%$.}
\label{fig:power-fermi}
\end{figure}

The dependence of the line-averaged, spectrally-inverted, Maxwellian-fit hot $T_e$ and hot $n_e$ are shown in Figure \ref{fig:power-fermi} as functions of RF power. Also shown is the normalized peak-to-peak floating potential of 200$\pm20$ MHz oscillations, $V_{pp}$, of a  Langmuir probe in the CC near the nozzle coil separating the CC from the FEC. Hot $T_e$ is seen to rise by a factor of 3 with power as does $V_{pp}$.  Auto-correlation and FFT analyses of $V_{pp}$ shows a narrow peak  at the RF  frequency, 19 MHz in this case, and decreasing amplitude narrow peaks at harmonics up to and beyond 200 MHz. The time-averaged Fourier amplitude of the 10th peak is 20 dB below that of the fundamental. The relative amplitudes of the harmonics change with time, and persist for 10s of $\mu$s before apparently changing mode. 

Our heuristic model makes use of the broadly coherent near-200-MHz oscillations, which are unrelated to the RF. 200 MHz is near the plasma frequency $\omega_{pe}$ of the warm population.

\begin{figure}[tbp]
\centering\includegraphics[scale=0.40]{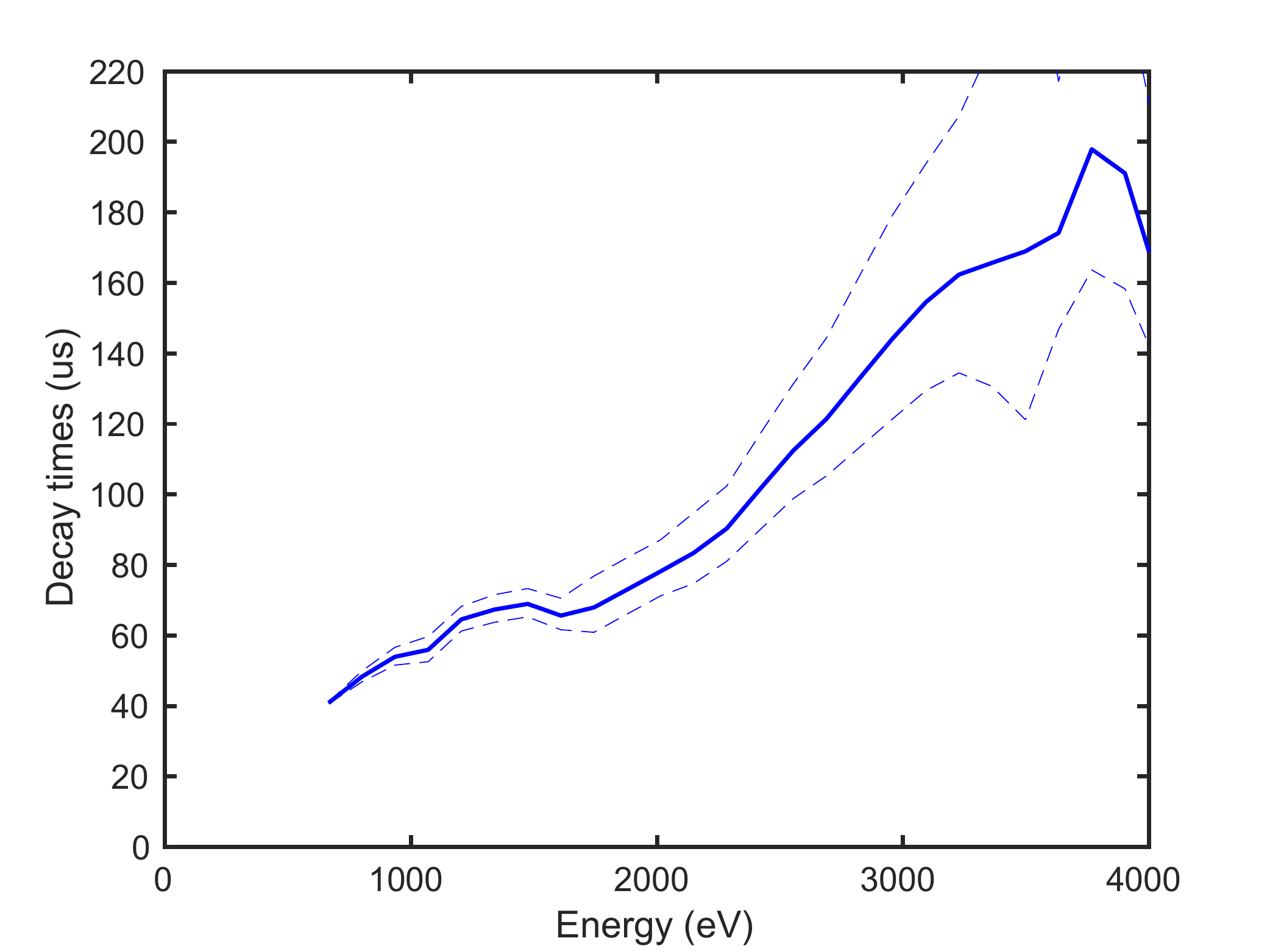}
\caption{Decay time of EEDF derived from X-ray spectrum, vs energy (solid line). 1$\sigma$ uncertainty of decay time (dashed lines).}
\label{fig:decay-fermi}
\end{figure}

By square-wave modulating the RF power, we can determine both the heating and loss-$\it{plus}$-cooling rates for these fast electrons. Figure \ref{fig:decay-fermi} (solid line) shows the decay time after cessation of the RF for the EEDF signal, using narrow bands of energies. Measurements made above 8 keV showed a continual rise of the decay time, though the error bars grow increasingly larger. Decay times longer than 1 ms have been seen for the highest energy measured. The rise time after the initiation of RF power also increases with X-ray energy. These findings indicate a gradual energization and loss mechanism, occurring over thousands of mirror transits.

\begin{figure}[tbp]
\centering\includegraphics[scale=0.82]{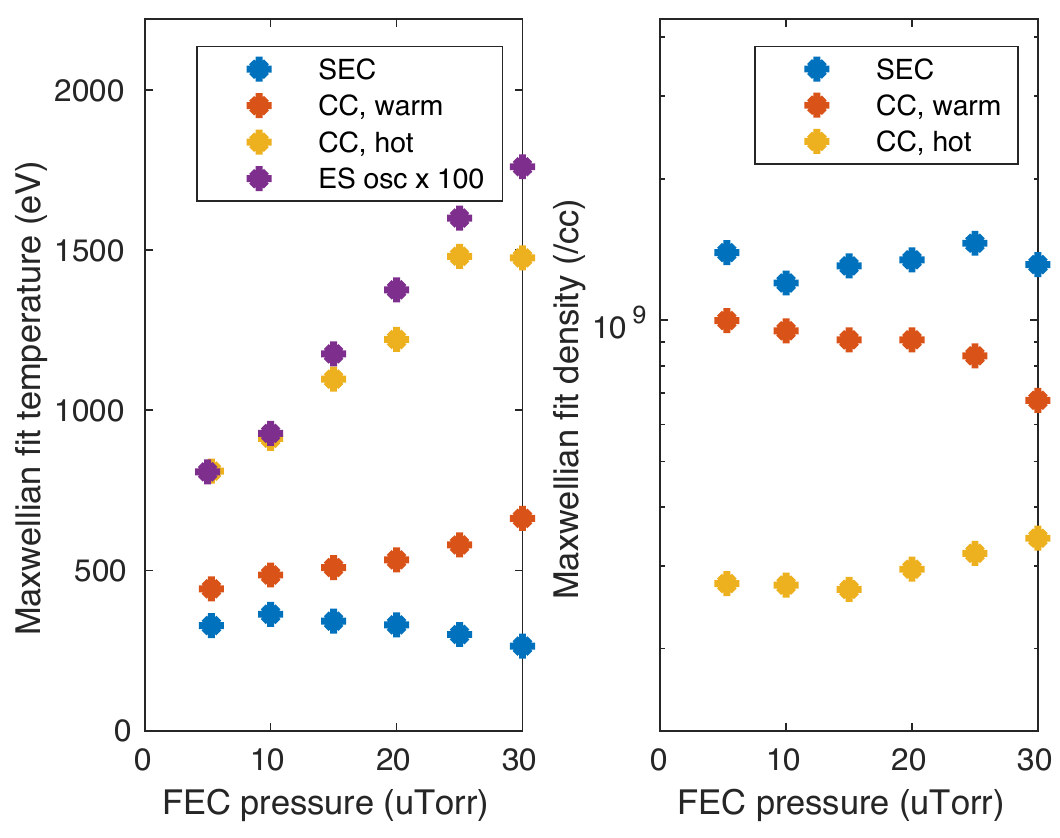}
\caption{Minority electron population $T_e$ and $n_e$ derived from X-ray signals and scaled peak-to-peak electrostatic oscillation (``ES osc'') $\it{vs}$ FEC gas pressure. The CC pressure was held constant at 0.133 mTorr of H$_2$ gas.}
\label{fig:FEC-press-fermi}
\end{figure}

\begin{figure}[tbp]
\centering\includegraphics[scale=0.6]{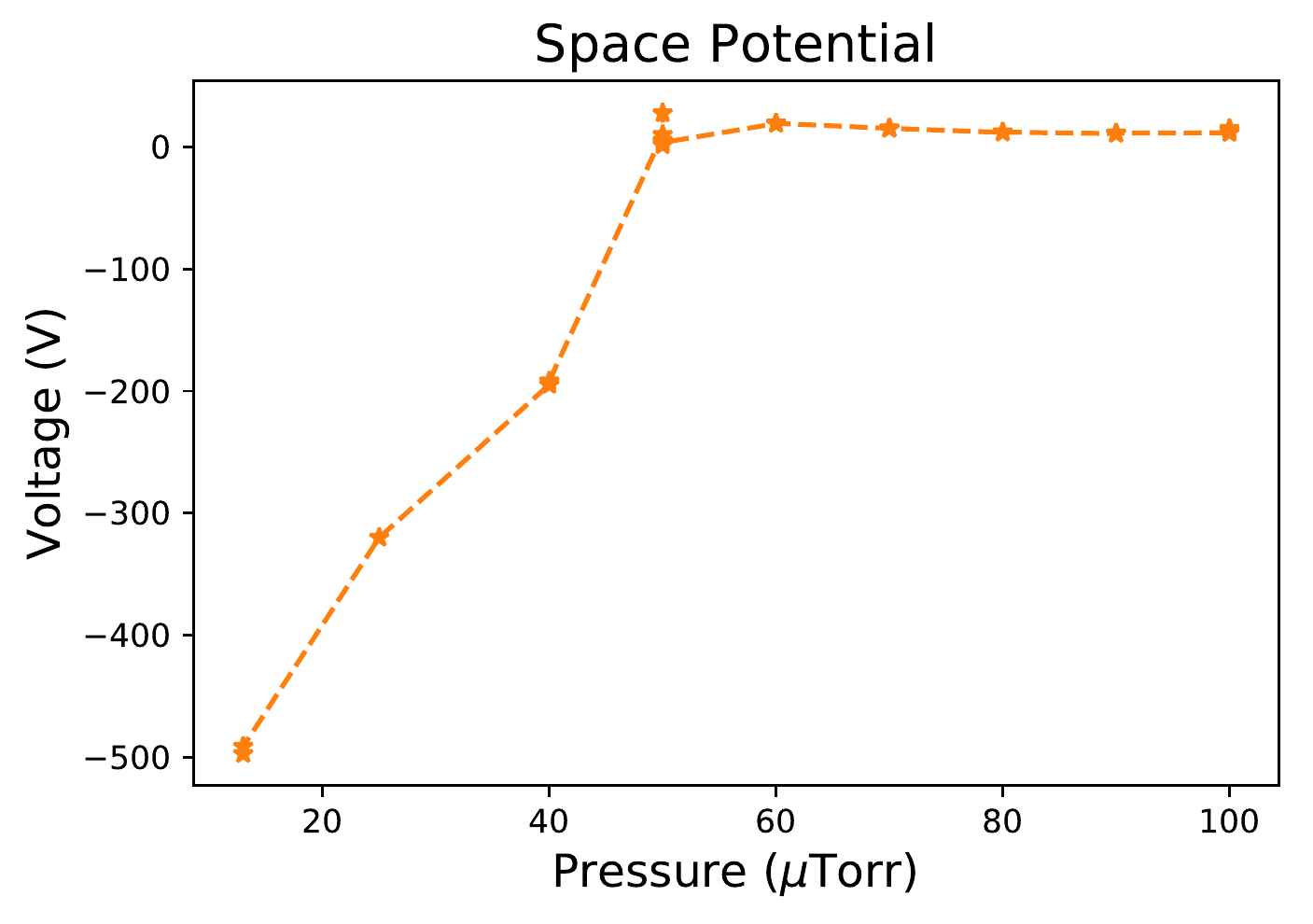}
\caption{FEC plasma space potential vs FEC gas pressure. }
\label{fig:FEC-potential-fermi}
\end{figure}

The dependence on FEC pressure of $V_{pp}$ and the CC and SEC effective temperatures and densities are shown in Figure \ref{fig:FEC-press-fermi} and the FEC space potential in \ref{fig:FEC-potential-fermi}. As the pressure is increased, the CC hot $T_e$ and $V_{pp}$ double while warm $T_e$ increases 30\% and SEC warm $T_e$ falls 30\%. Above p $\sim$ 50 $\mu$T, the hot population suddenly disappears simultaneously with the FEC space potential rising to 0 and the FEC end-plate (paddle, see Figure \ref{fig:schematic}) potential (not shown) also rises to zero from its highly negative potential, $\it{ca.}$ -1200 V. The (visible-wavelength) brightness and plasma density in the FEC and the heat flux to the paddle also rise dramatically as the gas pressure increases. The paddle potential is a measure of the relative fluxes of fast and bulk electrons. During these experiments, the plasma space potential in the CC and SEC is close to zero. These measurements broadly support the proposed mechanism for electrostatic oscillation generation and electron generation given later in this letter.

\begin{figure}[tbp]
\centering\includegraphics[scale=0.6]{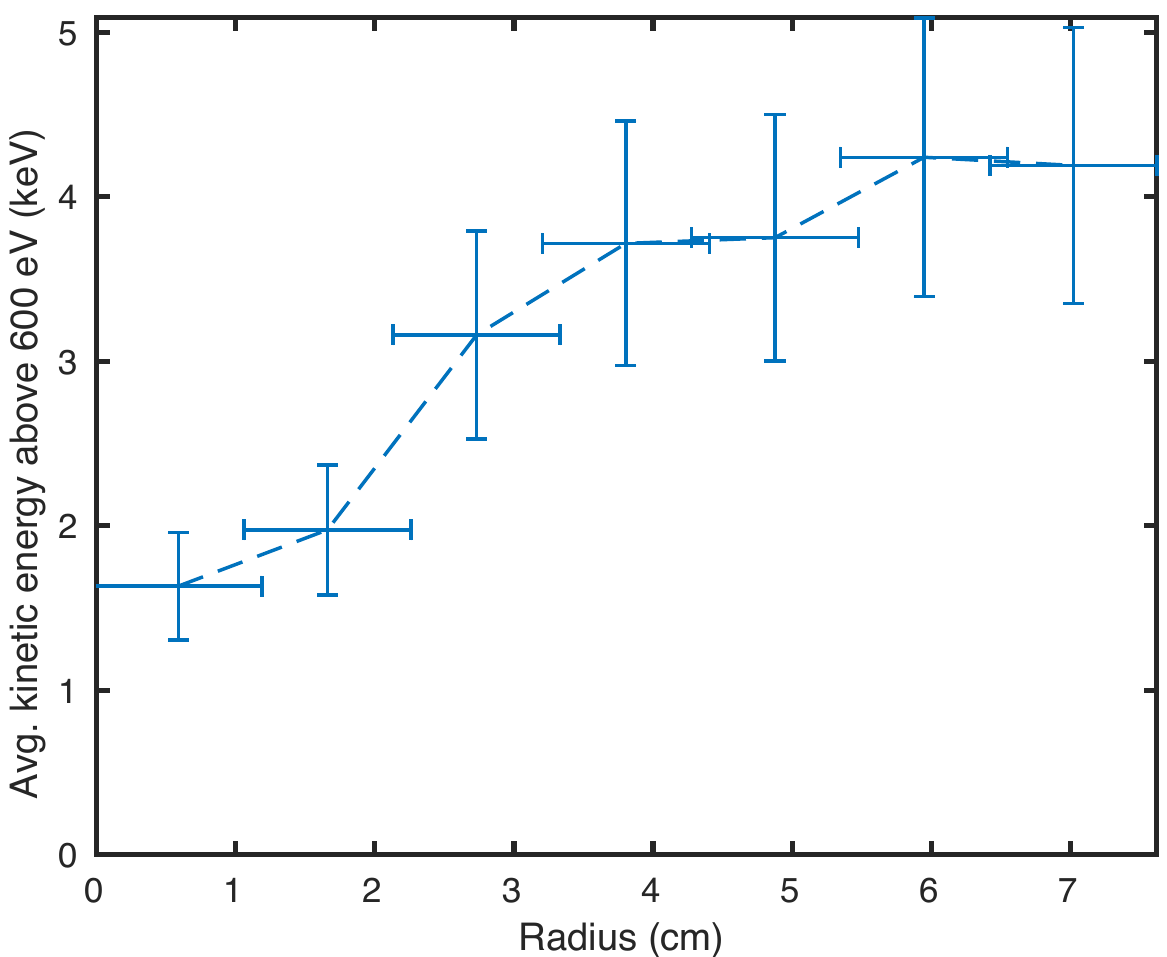}
\caption{The average electron kinetic energy for $E > 600$ eV, obtained by Abel inversion of deconvolved X-ray spectra as a function of radius.}
\label{fig:T_e-vs-r.pdf}
\end{figure}

The average electron kinetic energy is higher at larger radii, see Figure \ref{fig:T_e-vs-r.pdf}, one of several observations that strongly  supports the interpretation that $\mu$ non-conservation near z = 0 is the dominant cause of the required  velocity-space randomization needed for particle heating. 

The CC hot $T_e$ and maximum electron energy are both far greater than the paddle potential and potentials in the SEC. This informs us that electrostatic confinement is not the cause of the electron confinement or heating. 

We will now discuss our heuristic model of the process to which we attribute the presence of 3+ keV electrons.

At low FEC gas pressure, the space potential in the FEC is strongly negative compared to that in the central cell as seen in Figure \ref{fig:FEC-potential-fermi}. This is presumably because of the loss of 300 eV warm electrons from the CC to the FEC. Ionization occurs in the FEC by electron-impact ionization by these same 300 eV electrons. The electrons produced thereby will flow back into the CC as a beam with low temperature, $\sim$ 5 eV, and higher axially directed energy, $\sim$0.5 keV, similar to the electron beams seen in plasma double layers.\cite{schrittwieser_observation_1992} Higher gas pressures will create more ionization hence a stronger beam. At too high gas pressure, the higher bulk electron density dominates the space potential in the FEC, bringing it close to 0. Then a keV-scale beam back into the CC does not occur. 

The beam creates a two-stream instability as it enters the CC through the nozzle coil there, again as seen in double layers.\cite{schrittwieser_observation_1992} The estimated instability growth rate and energy decrement rate are such that the instability will saturate near the nozzle and the beam will fully dissipate. The two-stream creates the coherent $ \sim$ 200 MHz electrostatic oscillations of strength $\sim$ 50 Volts found there, as depicted (``ES osc'') in Figures \ref{fig:power-fermi} and \ref{fig:FEC-press-fermi}. This oscillation is an axial acceleration mechanism; an electron incident on the oscillation region may gain or lose on the order of 50 eV of energy, depending on the oscillation phase. The recurring bounces make this a Fermi acceleration process. The stronger 19 MHz signal does not contribute, as the voltage does not change significantly during an electron bounce. 

There are two apparent contradictions in applying this model to the data that we observe. Firstly, this mechanism would appear to increase the field-parallel energy of the electrons, decreasing their pitch angle and causing de-trapping. Secondly, the amplitude and frequency of the electrostatic oscillation would appear to limit the maximum electron energy to a value significantly less than 30 keV via adiabatic boundaries.\cite{lieberman_stochastic_1972}

The mechanism we propose to resolve both apparent contradictions has been known since the 1950's\cite{henrich_departure_1956,garren_individual_1958} and has been more thoroughly investigated since then.\cite{chirikov_stability_1978,tagare_motion_1986,zelenyi_quasiadiabatic_2013} That mechanism is $\mu$-non-conservation when passing through the z = 0 (midplane) of a mirror. This $\mu$ mobility is due to its motion through the steady-state curvature-changing vacuum field, not through interaction with fluctuations. Changes in $\mu$ can be of 0-th order, depending on a $\it{modern}$ adiabatic parameter that includes the parallel velocity and rate of field-curvature change.\cite{hastie_non-adiabatic_1969,cohen_nonadiabaticity_1978} 

This $\mu$-non-conservation is sufficient to maintain the trapped-ness of the particles, causing diffusion in $\mu$ between the passing/trapped value and a specific larger value of $\mu$ than predicted by the Chirikov criterion.\cite{tagare_motion_1986}

This $\mu$-non-conservation is also sufficient to solve the second apparent contradiction: Acceleration by this sinusoidal oscillation to a thermal-like distribution would appear to violate the adiabatic boundaries that a 2-dimensional (energy and oscillation phase) map would impose.\cite{lieberman_stochastic_1972} However, the evaluation of a four-dimensional (energy, oscillation phase, $\mu$, and gyrophase) map with coupling parameters relevant to the PFRC-II indicates that the size of the changes in $\mu$ is sufficient to destroy KAM surfaces and allow this acceleration.

This natural $\mu$ mobility serves the same purpose as the second ECRH frequency in two-frequency ECRH apparatus, and may have unknowingly existed at sufficient amplitudes in current and historical experiments.

The same $\mu$-non-conservation is responsible for an enhanced density of warm particles in the CC. X-ray measurements show that a portion of warm electrons become trapped in the CC and persist for thousands of transits, causing a 5$\times$ increase in density compared to the single-pass density expected from the warm electrons in the SEC. This is in general agreement with the stochastic region of phase space predicted by the Chirikov criterion to occur in every magnetic mirror.\cite{tagare_motion_1986}

Warm plasma from the SEC is necessary for the Fermi acceleration process. It overcomes the higher collision cross section of lower energy electrons and lowers  the FEC space potential to a strongly negative value. By energizing the electrons,  their rate of energy gain increases because of the shortened CC  transit time.

A second paper will discuss this model in detail, describing how a Maxwellian-like distribution is a natural result of the multi-dimensional Fermi acceleration.

$\bf{Acknowledgements}$ We are grateful for the contributions of T. Qian, P. Jandovitz and B. Berlinger.  This work was supported, in part, by the Program in Plasma Science and Technology, DOE contract DE-AC02-09CH11466, and IMOD Purchase Order Number 4440834795.


\bibliographystyle{aiaa}
\bibliography{references}

\begin{thebibliography}{10}
\newcommand{\enquote}[1]{``#1''}

\bibitem{malmberg_collisionless_1967}
Malmberg, J.~H. and Wharton, C.~B., \enquote{Collisionless {Damping} of
  {Large}-{Amplitude} {Plasma} {Waves},} {\em Physical Review Letters\/},
  Vol.~19, No.~14, Oct. 1967, pp.~775--778.

\bibitem{wharton_microwave_1968}
Wharton, C.~B. and Malmberg, J.~H., \enquote{Microwave {Scattering} from
  {Plasma} {Waves},} {\em The Physics of Fluids\/}, Vol.~11, No.~12, Dec. 1968,
  pp.~2655--2664.

\bibitem{yamada_study_1997}
Yamada, M., Ji, H., Hsu, S., Carter, T., Kulsrud, R., Bretz, N., Jobes, F.,
  Ono, Y., and Perkins, F., \enquote{Study of driven magnetic reconnection in a
  laboratory plasma,} {\em Physics of Plasmas\/}, Vol.~4, No.~5, May 1997,
  pp.~1936--1944.

\bibitem{lieberman_principles_2005}
Lieberman, M.~A. and Lichtenberg, A.~J., {\em Principles of {Plasma}
  {Discharges} and {Materials} {Processing}\/}, John Wiley \& Sons, April 2005,
  Google-Books-ID: m0iOga2XE5wC.

\bibitem{takahashi_high_2017}
Takahashi, K., Akahoshi, H., Charles, C., Boswell, R.~W., and Ando, A.,
  \enquote{High temperature electrons exhausted from rf plasma sources along a
  magnetic nozzle,} {\em Physics of Plasmas\/}, Vol.~24, No.~8, Aug. 2017,
  pp.~084503.

\bibitem{jandovitz_demonstration_2018}
Jandovitz, P., Swanson, C., Matteucci, J., Oliver, R., Pearcy, J., and Cohen,
  S.~A., \enquote{Demonstration of fast-electron populations in a low-pressure,
  low-power, magnetized {RF} plasma source,} {\em Physics of Plasmas\/},
  Vol.~25, No.~3, March 2018, pp.~030702.

\bibitem{alexeff_oscillations_1968}
Alexeff, I., Guest, G.~E., Montgomery, D., Neidigh, R.~V., and Rose, D.~J.,
  \enquote{Oscillations {Present} in {Plasma}-{Electron} {Heating} by an
  {Electron} {Beam},} {\em Physical Review Letters\/}, Vol.~21, No.~6, Aug.
  1968, pp.~344--347.

\bibitem{smullin_generation_1962}
Smullin, L.~D. and Getty, W.~D., \enquote{Generation of a {Hot}, {Dense}
  {Plasma} by a {Collective} {Beam}-{Plasma} {Interaction},} {\em Physical
  Review Letters\/}, Vol.~9, No.~1, July 1962, pp.~3--6.

\bibitem{demirkanov_plasma_1965}
Demirkanov, R.~A., Gevorkov, A.~F., and Popov, A.~F., \enquote{Plasma {Heating}
  under {Beam} {Instability} {Conditions},} {\em Plasma Physics and Controlled
  Nuclear Fusion Research. Proceedings of the Third International Conference on
  Plasma Physics and Controlled Nuclear Fusion Research\/}, 1965.

\bibitem{blinov_plasma_1967}
Blinov, P.~I., Zakatov, L.~P., Plakhov, A.~G., Chikin, R.~V., and Shapkin,
  V.~V., \enquote{Plasma {Heating} by an {Electric} {Beam} in a {Magnetic}
  {Mirror} {Machine},} {\em Soviet Physics JETP\/}, Vol.~25, No.~3, Sept. 1967,
  pp.~439.

\bibitem{arzhannikov_new_1988}
Arzhannikov, A.~V., Burdakov, A.~V., Kapitonov, V.~A., Koidan, V.~S.,
  Konyukhov, V.~V., Lebedev, S.~V., Mekler, K.~I., Nikolaev, V.~S., Postupaev,
  V.~V., Ryutov, D.~D., Shcheglov, M.~A., Sinitsky, S.~L., Voropaev, S.~G., and
  Vyacheslavov, L.~N., \enquote{New experimental results on beam-plasma
  interaction in solenoids,} {\em Plasma Physics and Controlled Fusion\/},
  Vol.~30, No.~11, 1988, pp.~1571.

\bibitem{howard_four-dimensional_1986}
Howard, J.~E., Lichtenberg, A.~J., Lieberman, M.~A., and Cohen, R.~H.,
  \enquote{Four-dimensional mapping model for two-frequency electron cyclotron
  resonance heating,} {\em Physica D: Nonlinear Phenomena\/}, Vol.~20, No.~2,
  June 1986, pp.~259--284.

\bibitem{lichtenberg_velocity_1986}
Lichtenberg, A.~J., Lieberman, M.~A., Howard, J.~E., and Cohen, R.~H.,
  \enquote{Velocity diffusion in two‐frequency electron cyclotron resonance
  heating,} {\em The Physics of Fluids\/}, Vol.~29, No.~4, April 1986,
  pp.~1061--1075.

\bibitem{bardet_hot-electron-plasma_1975}
Bardet, R., Briand, P., Dupas, L., Gormezano, C., and Melin, G.,
  \enquote{Hot-electron-plasma accumulation in the {CIRCE} mirror experiment,}
  {\em Nuclear Fusion\/}, Vol.~15, No.~5, 1975, pp.~865.

\bibitem{cohen_first_2012}
Cohen, S., \enquote{First operation of the {PFRC}-2 device,} {\em Bulletin of
  the {American} {Physical} {Society}\/}, Vol. Volume 57, Number 12, American
  Physical Society, Oct. 2012.

\bibitem{noauthor_amptek_nodate}
\enquote{Amptek – {X}-{Ray} {Detectors} and {Electronics},} .

\bibitem{swanson_using_2018}
Swanson, C., Jandovitz, P., and Cohen, S.~A., \enquote{Using
  {Poisson}-regularized inversion of {Bremsstrahlung} emission to extract full
  electron energy distribution functions from x-ray pulse-height detector
  data,} {\em AIP Advances\/}, Vol.~8, No.~2, Feb. 2018, pp.~025222.

\bibitem{schrittwieser_observation_1992}
Schrittwieser, R., Axnas, I., Carpenter, T., and Torven, S.,
  \enquote{Observation of double layers in a convergent magnetic field,} {\em
  IEEE Transactions on Plasma Science\/}, Vol.~20, No.~6, Dec. 1992,
  pp.~607--613.

\bibitem{lieberman_stochastic_1972}
Lieberman, M.~A. and Lichtenberg, A.~J., \enquote{Stochastic and {Adiabatic}
  {Behavior} of {Particles} {Accelerated} by {Periodic} {Forces},} {\em
  Physical Review A\/}, Vol.~5, No.~4, April 1972, pp.~1852--1866.

\bibitem{henrich_departure_1956}
Henrich, L.~R., \enquote{Departure of {Particle} {Orbits} from the {Adiabatic}
  {Approximation},} {\em Proc. {Conf}. on {Thermonuclear} {Reactions}\/},
  United States Atomic Energy Comission, Gatlinburg, Tennessee, 1956.

\bibitem{garren_individual_1958}
Garren, A.~A., Riddell, R.~J., Smith, L., Bing, G., Roberts, J.~E., Northrop,
  T.~G., and Henrich, L.~R., \enquote{Individual particle motion and the effect
  of scattering in an axially symmetric magnetic field,} {\em Journal of
  Nuclear Energy (1954)\/}, Vol.~7, No. 3-4, Sept. 1958, pp.~283--284.

\bibitem{chirikov_stability_1978}
Chirikov, B.~V., \enquote{Stability of the {Motion} of a {Charged} {Particle}
  in a {Magnetic} {Confinement} {System},} {\em Sov. J. Plasma Phys. (Engl.
  Transl.); (United States)\/}, Vol.~4:3, May 1978.

\bibitem{tagare_motion_1986}
Tagare, S.~G., \enquote{Motion of charged particles in an axisymmetric magnetic
  mirror,} {\em Physical Review A\/}, Vol.~34, No.~2, Aug. 1986,
  pp.~1587--1590.

\bibitem{zelenyi_quasiadiabatic_2013}
Zelenyi, L.~M., Neishtadt, A.~I., Artemyev, A.~V., Vainchtein, D.~L., and
  Malova, H.~V., \enquote{Quasiadiabatic dynamics of charged particles in a
  space plasma,} {\em Physics-Uspekhi\/}, Vol.~56, No.~4, 2013, pp.~347.

\bibitem{hastie_non-adiabatic_1969}
Hastie, R.~J., Hobbs, G.~D., and Taylor, J.~B., \enquote{Non-{Adiabatic}
  {Behaviour} of {Particles} in {Inhomogeneous} {Magnetic} {Fields},} {\em
  Plasma Physics and Controlled Nuclear Fusion Research. Proceedings of the
  Third International Conference on Plasma Physics and Controlled Nuclear
  Fusion Research. Vol. I\/}, 1969.

\bibitem{cohen_nonadiabaticity_1978}
Cohen, R.~H., Rowlands, G., and Foote, J.~H., \enquote{Nonadiabaticity in
  mirror machines,} {\em Physics of Fluids (1958-1988)\/}, Vol.~21, No.~4,
  April 1978, pp.~627--644.

\end{thebibliography}

\end{document}